\documentclass[12pt]{article}
\pdfoutput=1
\usepackage{jheppub}
\usepackage{amsmath}
\usepackage{amsfonts}
\usepackage{amssymb}
\usepackage{graphicx}

\usepackage[export]{adjustbox}
\newcommand{\bmat}{\left(\begin{array}}
\newcommand{\emat}{\end{array}\right)}

\def\yzero{\smash{\hbox{$y\kern-4pt\raise1pt\hbox{${}^\circ$}$}}}

\def\b{\beta}

\def\beq{\begin{equation}}
\def\eeq{\end{equation}}
\def\beqa{\begin{eqnarray}}
\def\eeqa{\end{eqnarray}}

\def\-{\hphantom{-}}

\def\s2{\frac{1}{\sqrt2}}

\def\beq{\begin{equation}}
\def\eeq{\end{equation}}
\def\beqa{\begin{eqnarray}}
\def\eeqa{\end{eqnarray}}

\def\IF{\relax{\rm I\kern-.18em F}}
\def\II{\relax{\rm I\kern-.18em I}}

\def\Dsl{\,\raise.15ex\hbox{/}\mkern-13.5mu D} 

\def\IS{{\bf {S}}}

\catcode`\@=11   
\newdimen\@rotdimen
\newbox\@rotbox  

\def\@vspec#1{\special{ps:#1}}
\def\@rotstart#1{\@vspec{gsave currentpoint currentpoint translate
   #1 neg exch neg exch translate}}
\def\@rotfinish{\@vspec{currentpoint grestore moveto}}
\def\@rotr#1{\@rotdimen=\ht#1\advance\@rotdimen by\dp#1%
   \hbox to\@rotdimen{\hskip\ht#1\vbox to\wd#1{\@rotstart{90 rotate}%
   \box#1\vss}\hss}\@rotfinish}
\def\@rotl#1{\@rotdimen=\ht#1\advance\@rotdimen by\dp#1%
   \hbox to\@rotdimen{\vbox to\wd#1{\vskip\wd#1\@rotstart{270 rotate}%
   \box#1\vss}\hss}\@rotfinish}%
\def\@rotu#1{\@rotdimen=\ht#1\advance\@rotdimen by\dp#1%
   \hbox to\wd#1{\hskip\wd#1\vbox to\@rotdimen{\vskip\@rotdimen
   \@rotstart{-1 dup scale}\box#1\vss}\hss}\@rotfinish}%
\def\@rotf#1{\hbox to\wd#1{\hskip\wd#1\@rotstart{-1 1 scale}%
   \box#1\hss}\@rotfinish}%
\def\rotate{\@ifnextchar[{\@rotate}{\@rotate[l]}}
\def\@rotate[#1]#2{\setbox\@rotbox=\hbox{#2}\@nameuse{@rot#1}\@rotbox}

\catcode`\@=12

\begin{document}

\makeatletter
\@addtoreset{equation}{section}
\makeatother
\renewcommand{\theequation}{\thesection.\arabic{equation}}
\pagestyle{empty}
\vspace*{0.5in}
\vspace{1.5cm}
\begin{center}
\Large{\bf Dynamical Cobordism and the Beginning of Time:\\
Supercritical Strings and Tachyon Condensation
}
\\[8mm] 

\large{Roberta Angius, Matilda Delgado,  Angel M. Uranga\\[4mm]}
\footnotesize{Instituto de F\'{\i}sica Te\'orica IFT-UAM/CSIC,\\[-0.3em] 
C/ Nicol\'as Cabrera 13-15, 
Campus de Cantoblanco, 28049 Madrid, Spain}\\ 
\footnotesize{\href{roberta.angius@csic.es}{roberta.angius@csic.es},  \href{mailto:matilda.delgado@uam.es}{matilda.delgado@uam.es},  \href{mailto:angel.uranga@csic.es}{angel.uranga@csic.es}}

\vspace*{10mm}

\small{\bf Abstract} \\
\end{center}
\begin{center}
\begin{minipage}[h]{\textwidth}
We describe timelike linear  dilaton backgrounds of supercritical string theories as time-dependent Dynamical Cobordisms in string theory, with their spacelike singularity as a boundary defining the beginning of time. We propose and provide compelling evidence that its microscopic interpretation corresponds to a region of (a strong coupling version of) closed tachyon condensation. We argue that this beginning of time is closely related to (and shares the same scaling behaviour as) the bubbles of nothing obtained in a weakly coupled background with lightlike tachyon condensation. As an intermediate result, we also provide the description of the latter as lightlike Dynamical Cobordism.
\end{minipage}
\end{center}
\newpage
\setcounter{page}{1}
\pagestyle{plain}
\renewcommand{\thefootnote}{\arabic{footnote}}
\setcounter{footnote}{0}

\tableofcontents

\vspace*{1cm}

\newpage

\section{Introduction}

One of the outstanding questions in string theory is the understanding of time-dependent backgrounds and in particular the resolution of cosmological (i.e. spacelike) singularities (see \cite{Quevedo:2002xw,Craps:2006yb} for reviews). On general grounds, and in analogy with timelike singularities, one may expect that stringy effects smooth out the singularity, thus providing a microscopic description of the beginning of time. 

This is a natural proposal from the perspective of the Swampland Cobordism Conjecture \cite{McNamara:2019rup}, which states that any consistent theory of quantum gravity should admit configurations ending spacetime, namely boundaries or general cobordism defects leading to walls of nothing\footnote{A prototype is the infinite volume limit of bubbles of nothing in higher-dimensional compactifications in which the internal space shrinks to zero size \cite{Witten:1981gj} (see \cite{Ooguri:2017njy,GarciaEtxebarria:2020xsr,Dibitetto:2020csn,Bomans:2021ara} for some recent work).}. This also resonates with (a Lorentzian version of) the no-boundary proposal for the Hartle-Hawking wavefunction of the universe \cite{Hartle:1983ai}.

From this perspective, such cosmological solutions would correspond to dynamical time-dependent configurations with a beginning of time given by a cobordism defect extending in the spatial directions. This appealing picture is however hampered by the general lack of understanding of the microscopic structure of spacelike singularities.

The cobordism conjecture has been exploited at the topological level with interesting results, see e.g. \cite{GarciaEtxebarria:2020xsr,Ooguri:2020sua,Montero:2020icj,Dierigl:2020lai,Hamada:2021bbz,Blumenhagen:2021nmi,Andriot:2022mri}. On the other hand, there is substantial progress in understanding the implications of the cobordism conjecture at the dynamical level. The configurations dubbed Dynamical Cobordisms in \cite{Buratti:2021yia,Buratti:2021fiv,Angius:2022aeq} (see also \cite{Blumenhagen:2022mqw})\footnote{For the related topic of solutions in theories with dynamical tadpoles, see \cite{Dudas:2000ff,Blumenhagen:2000dc,Dudas:2002dg,Dudas:2004nd} for early work and \cite{Basile:2018irz,Antonelli:2019nar,Mininno:2020sdb,Basile:2020xwi,Basile:2021mkd,Mourad:2022loy} for related recent developments.} describe spacetime dependent solutions in which the fields run until they hit a real-codimension 1 singularity at finite distance in spacetime, at which certain scalars run off to infinite distance in field space. In several examples of such spatially varying solutions, the timelike singularities had a known string theory UV description, which displayed an end of spacetime. Remarkably, \cite{Angius:2022aeq} showed that in the effective theory description these singularities (dubbed end-of-the-world (ETW) branes) follow universal scaling laws, and are characterized by a single critical exponent.

In this paper we take the natural next step of starting the study of time-dependent Dynamical Cobordism with spacelike singularities and of shedding some light on their resolution. The particular arena to explore these ideas are timelike linear dilaton backgrounds in supercritical bosonic string theory. 

Supercritical string theories provide consistent versions of string theory in a general number $D$ of spacetime dimensions, provided a suitable timelike linear dilaton background is turned on \cite{Myers:1987fv,Antoniadis:1988aa,Antoniadis:1988vi,Antoniadis:1990uu}. They provide an excellent testing ground for general features of string theory (see \cite{Chen:2021emg} for a recent example). In particular, and as will be relevant to our discussion, they constitute a setup in which closed string tachyon physics has been subject to quantitative analysis (see e.g. \cite{Hellerman:2006nx,Hellerman:2006ff,Hellerman:2007fc,Swanson:2008dt,Berasaluce-Gonzalez:2013sna,Garcia-Etxebarria:2014txa,Garcia-Etxebarria:2015ota})\footnote{See \cite{Adams:2001sv,Uranga:2002ag,DeAlwis:2002kp} and references therein for discussion of the fate of localized closed tachyons and related instabilities.}. For our purposes, the main property of these theories is that the timelike linear dilaton background makes them one of the simplest time-dependent setups in string theory.

We express these backgrounds as time-dependent Dynamical Cobordisms, exhibiting their beginning of time singularity and characterizing it as an ETW brane, with a precise critical exponent. We moreover propose, providing non-trivial support for it, that the stringy resolution of the singularity involves a region of (the strong coupling version of) bulk tachyon condensation. This is a realization of the mechanism in \cite{McGreevy:2005ci} in a different setup which, as promised, provides a stringy analogue of the Hartle-Hawking proposal.

Our approach is based on the realization that the beginning of time singularity, and the walls of nothing described via lightlike tachyon condensation in \cite{Hellerman:2006nx} (see also \cite{Hellerman:2006ff,Hellerman:2006hf,Hellerman:2007ym,Hellerman:2007zz, Hellerman:2007fc,Swanson:2008dt} for related results) admit an ETW brane description in the effective theory with exactly the same critical exponent. Moreover, we show that these configurations, which seemingly contain two intersecting ETW walls, actually contain a single recombined one with two different asymptotic regions, a lightlike one corresponding to tachyon condensation at weak coupling and a spacelike one at strong coupling corresponding to the beginning of time.

A potential caveat to our analysis is the use of effective theories to describe tachyon condensation phenomena, which involve stringy scales and are not fully understood for closed tachyons (see \cite{Yang:2005rx} and references therein for further discussion). We however encounter that the main feature of the ETW wall, the critical exponent, is surprisingly robust under corrections of the effective action. This suggest that the main results may survive beyond the validity of the tools used to extract it in the present work. The same considerations apply to the study of the beginning of time singularity, which lies at strong coupling.

The paper is organized as follows. In Section \ref{sec:dyncob} we recall the Dynamical Cobordisms of \cite{Buratti:2021yia,Buratti:2021fiv,Angius:2022aeq}, and the structure of the ETW branes in terms of their critical exponent. In Section \ref{sec:time-dependent} we discuss the timelike linear dilaton background as a time-dependent solution: In Section \ref{sec:linear-dilaton} we express it as a Dynamical Cobordism with a beginning of time; in Section \ref{sec:etw-bot} we describe the singularity at the beginning of time as an ETW brane; and in Section \ref{sec:timelike-tachyon} we explore its UV description in terms of a timelike tachyon condensate. In Section 
\ref{sec:lightlike-tachyon} we discuss walls of nothing arising in lightlike tachyon condensation and show that they correspond to Dynamical Cobordisms with a lightlike ETW brane: in Section \ref{sec:lightlike-worldsheet} we recall the worldsheet description, and in Section \ref{sec:lightlike-spacetime} we provide their spacetime description and characterize their ETW brane and critical exponent. In Section \ref{sec:recombination} we combine results and formulate our proposal that the UV description of the beginning of time in the linear dilaton background is (a strong coupling version of) closed tachyon condensation. In Section \ref{sec:conclusions} we offer some final thoughts. In Appendix \ref{app:quench} we mention that the dimension quenching mechanism in \cite{Hellerman:2006ff,Hellerman:2007fc} can be described as a dynamical cobordism  describing an interpolating wall \cite{Buratti:2021fiv} between theories of different dimension. Some calculational details have been postponed to Appendices \ref{app:partitionfunction}, \ref{app:scalingrelation}.

\section{Overview of Dynamical Cobordisms}
\label{sec:dyncob}

In a series of papers \cite{Buratti:2021yia,Buratti:2021fiv,Angius:2022aeq} the analysis of dynamical spacetime-dependent solutions realizing cobordisms to nothing was initiated  (see also \cite{Blumenhagen:2022mqw}). Such solutions, from the perspective of the lower-dimensional effective field theory, present universal features that allow them to be described in a general framework as follows; 

Consider the lower-dimensional EFT to be $D$-dimensional\footnote{We use $D$ for the spacetime dimension and ${\cal D}$ for the field space distance.} Einstein gravity coupled to a scalar with arbitrary potential (in $M_{Pl}=1$ units):
\begin{equation}
	S = \int d^{D}x\, \sqrt{-g}\,  \left( \frac 12 R - \frac{1}{2} \left( \partial \phi\right)^{2} - V(\phi) \right) \,.
	\label{ddim-action}
\end{equation}
We consider solutions in which the metric and scalar vary along one coordinate, denoted by $y$. The ansatz for the metric is
\beqa
ds^2=e^{-2\sigma(y)}ds_{D-1}^{\, 2} + dy^2\, .
\label{dyncob-metric-ansatz}
\eeqa
Here we follow earlier references and considered space-dependent running solutions. The sign flips necessary for time-dependent ones will be taken into account in the concrete examples of later sections.

In the following we take a flat metric for the $D-1$ dimensional slices. All solutions that describe a cobordism to nothing present a spacetime singularity at finite spacetime distance $\Delta$ where the scalars explore an infinite distance $\mathcal{D}$ in field space, this is the location of the ETW wall. Indeed, the solution does not extend beyond this point which, without loss of generality, we choose to be $y=0$. 

One of the highlights of the analysis in \cite{Angius:2022aeq} is that the solutions near ETW branes behave in a simple way. We quote some of the main expressions encapsulating this
\begin{equation} 
\label{scalings-phi-sigma-R}
	\phi(y) \simeq - \frac{2}{\delta} \log y \quad, \quad \sigma(y) \simeq - \frac{4}{(D-2)\delta^{2}} \,\log y \, \quad ,\quad |R|   \simeq \frac{1}{y^2} \, .
\end{equation}
with $\delta$ a scaling coefficient which characterizes the local solution near the ETW brane, and $|R|$ is the spacetime scalar curvature. Although \cite{Angius:2022aeq} focused on space-dependent running solutions, it is straightforward to extend the discussion to time-dependent ones, and recover the same scaling laws. 

From the above profiles, all solutions describing ETW walls present universal scaling relations between $\Delta$, ${\cal D}$ and the spacetime scalar curvature $|R|$, as follows
\beqa
\Delta \sim e^{-\frac {\delta}{2} {\cal D}} \quad, \quad |R|\sim e^{\delta{\cal D}}\, .
\label{etw-scalings}
\eeqa

We also get that the scalar potential behaves as\footnote{Note that if $y$ is timelike, the overall sign of the potential changes. The quantity $c$ is a positive constant related to the boundary condition used when solving the equations of motion. Subleading corrections to the potential can be included by promoting $c$ to a function $c=c(\phi)$ with slower growth than an exponential. }
\begin{equation} 
\label{eq:sol-V}
	V(\phi) \simeq - a\,c \, e^{\delta\, \phi} \, ,
	\end{equation}
for a constant $a<1$ related to $\delta$ by
\begin{equation} \label{eq:gamma}
	\delta = 2 \sqrt{\frac{D-1}{D-2}\left( 1-a\right)} \, .
\end{equation}

\section{Supercritical strings as time-dependent Dynamical Cobordism}
\label{sec:time-dependent}

In this section we discuss the maximally symmetric configuration of supercritial strings, and interpret the necessary linear dilaton background as a running solution which satisfies the properties of a time-dependent Dynamical Cobordism. The local behaviour is hence that of an ETW brane.

In this work we focus on the supercritical bosonic theory. We expect similar ideas to apply to other supercritical theories, including supercritical type 0 or heterotic superstrings \cite{Hellerman:2006ff,Hellerman:2007fc}. 

\subsection{Linear timelike dilaton as Dynamical Cobordism}
\label{sec:linear-dilaton}

Consider bosonic string theory in $D$-dimensional Minkowski space (in the string frame). In order to satisfy the central charge constraint for the theory, there is a linear dilaton background
\beqa
\Phi=v_MX^M\, ,
\eeqa
with
\beqa
v\cdot v =-\frac{D-26}{6\alpha'}\, ,
\label{central-charge-condition}
\eeqa
with contractions defined with respect to the flat Minkowski metric.

Hence, supercritical strings require a timelike dilaton gradient, whereas subcritical strings require an spacelike one. The critical $D=26$ theory does not require a dilaton profile for consistency, but does admit a lightlike dilaton background. We thus expect our discussion to extend this background of the critical theory as well.

These linear dilaton theories define conformal theories exactly in $\alpha'$, which implies that they satisfy the equations of motion of the spacetime (string frame) action
\beqa
S_{\rm str.}=\frac {1}{2}\int d^Dx \sqrt{-G_{(s)}}e^{-2\Phi} \left[-\frac{2(D-26)}{3\alpha'}+R_{(s)}+4(\partial \Phi)^2
\right]\, .
\eeqa
We explicitly denote string frame quantities with an $s$ subindex, while quantities with no subindex are implicitly defined in the Einstein frame.

In the following we focus on supercritical strings and timelike dilaton background 
\beqa
\Phi=-q X^0\, ,
\eeqa
where $q\equiv v^0$. Here we have absorbed a possible additive constant by shifting time, so that the dilaton vanishes at $X^0=0$. From the two solutions of \eqref{central-charge-condition} we choose the one leading to weak coupling $g_s=e^\Phi$ in the future $X^0\to\infty$, namely
\beqa
q = \sqrt{\frac{D-26}{6\alpha'}} \ .
\label{defq}
\eeqa
In the next section, we reinterpret this linear dilaton background as a running solution with an ETW wall at the origin of time. 

\subsection{The ETW brane at the beginning of time}
\label{sec:etw-bot}

The spacetime physics of the singularity was considered from a cosmological perspective in \cite{Hellerman:2006nx}. Here we instead study it from the perspective of the ETW branes of dynamical cobordism in section \ref{sec:dyncob}.

To discuss the spacetime physics, we focus on the Einstein frame, so the metric reads
\beqa
ds^{\, 2}=\exp\Big(\,{\frac{4qX^0}{D-2}}\,\Big)\,\eta_{MN} dx^Mdx^N\, .
\label{linear-dilaton-einstein-metric1}
\eeqa
We see that at $X^0\to -\infty$ the warp factor goes to zero, and we hit a singularity. We can introduce a time coordinate $y$ giving the invariant interval to the singularity as
\beqa
y=\frac{D-2}{2q} \exp\Big(\frac{2qX^0}{D-2}\Big)\, ,
\eeqa
in terms of which \eqref{linear-dilaton-einstein-metric1} is recast as the time-dependent version of \eqref{dyncob-metric-ansatz}:
\beqa
ds^2=-dy^2+\frac{4q^2y^2}{(D-2)^2} \, dx^mdx^m\, ,
\eeqa
 for $m=1,\ldots, D-1$. We thus obtain
 \beqa
 \sigma(y)=-\log y \, .
 \eeqa
Comparing this to \eqref{scalings-phi-sigma-R} gives:
\beqa
\delta=\frac{2}{\sqrt{D-2}}\, .
\label{linear-dilaton-delta}
\eeqa

Expressing the dilaton in terms of a scalar $\phi$ with canonical kinetic term in the Einstein frame as
\beqa \label{eq:normalized-dil}
\phi=\frac {2}{\sqrt{D-2}} \,\Phi\sim -\sqrt{D-2}\,\log y\, .
\eeqa
This is precisely the scaling relation for the scalar \eqref{scalings-phi-sigma-R} for the value of $\delta$ in \eqref{linear-dilaton-delta}.

We also get the expected scaling of the potential. The Einstein frame action gives
\beqa
S=\frac 12\int d^Dx \sqrt{-G}\,\Big[\,R\,-\frac{4}{D-2}\,(\partial \Phi)^2\,-\,\frac{2(D-26)}{3\alpha'}\exp\Big(\,\frac{4\Phi}{D-2}
\,\Big)
\,\Big]\, ,
\eeqa
which, comparing with \eqref{eq:sol-V} and using the normalized dilaton \eqref{eq:normalized-dil} yields the precise value of $\delta$ in \eqref{linear-dilaton-delta}.

To conclude, we recover the scaling relations \eqref{etw-scalings}, which state that the configuration hits an ETW singularity at finite time in the past at which the scalar runs off to infinite distance in field space. According to the cobordism interpretation of such singularities in \cite{Buratti:2021fiv,Angius:2022aeq}, it defines a beginning of time, a boundary in the time direction, for this solution.

 The microscopic description of the ETW brane requires some understanding of spacelike defects in string theory, which remains mostly {\em terra incognita}\footnote{See e.g. \cite{Gutperle:2002ai} for attemps invoking S-branes)}. 
 In our particular example, this is even more so since it lies at strong coupling\footnote{For the critical type IIA with a lightlike dilaton background, a microscopic description for the analogous singularity was proposed in \cite{Craps:2005wd}, based on M(atrix) theory \cite{Banks:1996vh}. Such a description does not seem feasible in our case.}. 
Despite these difficulties, we find compelling evidence that the microscopic description of our spacelike ETW brane is the strong coupling avatar of tachyon condensation. We propose a direct approach to this proposal in the next section, and a further indirect, but quantitatively more reliable, route to support this picture in Section \ref{sec:lightlike-tachyon}.

\subsection{The timelike tachyon case}
\label{sec:timelike-tachyon}

The resolution of spacelike singularities in a tractable worldsheet approach was addressed in \cite{McGreevy:2005ci} in a setup with a shrinking 1-cycle, in terms of the condensation of a closed string tachyon in the winding sector (see \cite{McGreevy:2006hk,Green:2007tr} for proposed higher-genus generalizations). In short, the regime near the singularity was proposed to be coated by a longer duration region in which the tachyon condenses with an exponential profile.  The latter describes an effective Liouville wall in the time direction, beyond which no string excitation can propagate. This was argued to be a stringy definition of the {\em nothing} in the Hartle-Hawking description of the wavefunction of the universe \cite{Hartle:1983ai}. In this picture, spacetime emerges smoothly as the tachyon turns off. 
In our terms, it describes a cobordism to nothing in the time direction.

In this section we explore a similar interpretation for the spacelike singularity encountered in our timelike linear dilaton setup. The idea is to consider an exponential profile for the closed string tachyon of supercritical bosonic theory. The tachyon couples to the worldsheet as a 2d potential. The condition for this deformation to be marginal, to linear order in conformal perturbation theory, or equivalently, the linearized spacetime equation of motion for the tachyon is
\beqa
\partial^2 {T}(X) - 2\, v^M\, \partial_M {T}(X) + \frac{4}{\alpha'} {T}(X) = 0\ .
\label{tachyon-eom}
\eeqa
We will discuss corrections to this later on.

For a  general tachyon exponential profile
\beqa
T(X^M) = \mu\, \exp( \beta_M \, X^M ) \ .  
\label{tachyon-profile}
\eeqa
we obtain a condition on $\beta$:
\beqa
\beta\cdot\beta -2v\cdot\beta+\frac{4}{\alpha'}=0\, .
\eeqa

We now focus on a timelike tachyon profile
\beqa
T=\mu\exp (-\beta^0X^0)\, ,
\eeqa
with the condition
\beqa
-(\beta^0)^2+2q\beta^0+\frac{4}{\alpha'}=0\, .
\eeqa
There are two solutions to this quadratic equation. A possibility is to choose $\beta^0<0$, so that the tachyon grows for late times $X^0\to\infty$. This is a good strategy to study the process of tachyon condensation in a weakly coupled regime, see e.g. \cite{Swanson:2008dt}. In fact, it is closely related to our approach (albeit for lightlike tachyons) in Section \ref{sec:lightlike-tachyon}. 

Here, instead, we are interested in having tachyon condensation at the beginning of time, to provide a resolution of the spacelike ETW brane at $y=0$, hence we need the tachyon to grow in the past $X^0\to -\infty$, we thus require $\beta^0>0$. Using \eqref{defq} we have
\beqa
\beta^0=\frac{\sqrt{D-26}+\sqrt{D-2}}{\sqrt{6\alpha'}}\, .
\eeqa

We may now compare the relative growth of the string coupling $g_s=e^\Phi$ and of the tachyon as $X^0\to-\infty$ to assess if the tachyon condensation could be studied using worldsheet techniques. We have
\beqa
T/g_s=\mu\exp\Big( -\sqrt{\frac{D-2}{6\alpha'}} X^0\Big) \, .
\label{hierarchy}
\eeqa
This shows that the tachyon grows parametrically faster than the string coupling as $X^0\to-\infty$. This leads to the expectation that the worldsheet analysis provides a reliable description of the physics at early times. In analogy with \cite{McGreevy:2005ci}, and based on the extensive analysis in \cite{Hellerman:2006nx,Hellerman:2006ff,Hellerman:2007fc}, the presence of the worldsheet potential creates a Liouville wall expelling all string excitations, providing a microscopic definition of an ETW brane in time.

The drawback of this approach is that it relied on trusting the linearized deformation approximation, which is expected to experience strong higher order corrections\footnote{For the $\beta^0<0$ solution, corrections are interestingly expected to be suppressed in a large $D$ approximation, as exploited in \cite{Hellerman:2004qa}. Although large $D$ could be interesting in our $\beta^0>0$ case to increase the hierarchy in \eqref{hierarchy}, it does not lead to a similar suppression of corrections.}. Therefore the scenario can be at most regarded as a qualitative description. In the next Section we turn to a different approach, involving $\alpha'$ exact solutions.

\section{Lightlike tachyon condensation}
\label{sec:lightlike-tachyon}

We are thus led to consider solutions under better control.
In this Section we consider an $\alpha'$-exact solution of the supercritical linear dilaton theory with tachyon profile along a lightlike direction. As established in \cite{Hellerman:2006nx,Hellerman:2006ff} for the bosonic theory, at late times this leads to a wall of nothing moving at the speed of light, analogous to the asymptotic behaviour of a bubble of nothing. After recalling the argument, we carry out a new spacetime analysis that shows that at late times the background corresponds to a lightlike ETW brane, and show that its critical exponent is exactly the same as for the beginning of time ETW brane of the previous section. This tantalizing relation is a strong support for our interpretion of the beginning of time is (a strongly coupled version of) a closed string tachyon condensation phase, discussed in Section \ref{sec:recombination}.

\subsection{Lightlike tachyon in the worldsheet description}
\label{sec:lightlike-worldsheet}

Consider introducing an exponential tachyon background \eqref{tachyon-profile} along a lightlike direction $X^+=(X^0+X^1)/\sqrt{2}$
\begin{equation}
    T=\mu\exp(\beta X^+)\, .
    \label{lightlike-tachyon-profile}
\end{equation}

The linearized tachyon marginality condition \eqref{tachyon-eom} is satisfied for
\begin{equation}
    \beta=\frac{2\sqrt{2}}{q\alpha'}\, .
\end{equation}

At late times $X^0\to\infty$, the string coupling is small and one may perform a reliable worldsheet analysis.
As shown in \cite{Hellerman:2006nx} and contrary to the timelike tachyon case, the deformation by the operator \eqref{lightlike-tachyon-profile} is exact, as higher order corrections in the perturbation vanish, since the lightlike nature of the insertions prevent the existence of non-trivial Wick contractions. 
Furthermore, in light-cone coordinates the propagator of the $X^{+/-}$ fields is oriented from $X^{+}$ to $X^{-}$ and we know that all interaction vertices introduced by the tachyon potential only depend on $X^{+}$. These two facts combined show that there are no possible Feynman diagrams beyond tree-level, which implies the solution is exact in $\alpha'$. One can thus conclude that the linearized tree-level solution \eqref{lightlike-tachyon-profile} is exactly conformally invariant.

The tachyon couples as a worldsheet potential, which grows infinitely at $X^+\to \infty$. This 2d potential prevents any string modes from entering the corresponding region, which thus becomes a region of nothing. The physical interpretation of this is that the tachyon configuration describes a wall of nothing propagating at the speed of light, which effectively ends spacetime at an effective value of $X^+$. 

The finite range in $X^+$ can be estimated by e.g. cutting off $X^+$ when the $T=1$. This gives 
\beqa
\Delta X^+=-\beta^{-1}\log \mu/\mu^*\, ,
\label{range-x+}
\eeqa 
where $\mu^*$ defines a reference position from which we measure the range to the wall. A more precise derivation follows from the {\em gedanken} experiment of solving the motion of classical strings incoming into the tachyon wall \cite{Hellerman:2006nx}. The initial speed reduces to zero at a turning point, after which the string is pushed back by the tachyon wall and its speed asymptotes to that of light. The turning point position in $X^+$ in the formulas in \cite{Hellerman:2006nx} gives back the result \eqref{range-x+}.

Given the importance of the notion of finiteness on the location of the tachyon wall, we provide an alternative derivation, carried out by adapting the techniques in \cite{McGreevy:2005ci}. We briefly sketch the results here and give more computational details in Appendix \ref{app:partitionfunction}. Decomposing the field $X^+(\tau, \sigma)$ into its zero and nonzero modes $X^+ (\tau, \sigma) = X^+_0 + \widehat{X}^+(\tau, \sigma)$ and performing a Wick rotation, the Euclidean partition function reads:
\begin{equation}\label{eq:deformedpartfunc}
Z \left( \mu \right) = \int dX^+_0 \int \mathcal{D} \widehat{X}^+ \mathcal{D} X^- \mathcal{D}X^i \mathcal{D}g \mathcal{D}({ ghosts}) e^{- S^{deformed}_{E}}\, ,
\end{equation}
with the Euclidean action: 
\begin{equation}
\begin{split}\label{EuclidianAction2}
    S_E^{\rm. deformed} = & \frac{1}{2 \pi \alpha'} \int d^2 \sigma_E \sqrt{g} \left[\partial_{\sigma^0} \widehat{X}^+ \partial_{\sigma^0} X^- + \partial_{\sigma^1} X^- \partial_{\sigma^1} \widehat{X}^+ +  \partial_{\alpha} X^i \partial_{\alpha} X^i \right] \\  & + \frac{1}{2 \pi} \int d^2 \sigma_E \sqrt{g} R_2 \Phi (X) + \frac{i \mu_E}{2 \pi} \int d^2 \sigma_E \sqrt{g} e^{\beta X^+} \, .\\ \end{split}\end{equation}
From this we see that when the tachyon condenses, at large $X^{+}$, the path integral becomes suppressed. This results in a truncation of contributions to the integral coming from string oscillations with $X^+ \mapsto \infty$.  This is the same mechanism as that of a Liouville wall in Liouville theory: no physical degrees of freedom exist in this region. In fact, one can show that the partition function in \eqref{eq:deformedpartfunc} can be directly related to that of the free theory (with no tachyon deformation) as follows.

After integrating out the zero-mode $X^{+}_0$, one can show that:
\begin{equation}
      \frac{\partial Z}{\partial \mu_E} = -\frac{1}{\beta \mu_E} \int \mathcal{D} \widehat{X}^+ \mathcal{D} X^- \mathcal{D}X^i \mathcal{D}g \mathcal{D}(ghosts) e^{-S_E^{free}} \, ,
\end{equation}
where $S_E^{\rm free}$ is the euclidean action of the worldsheet theory without the tachyon potential. Integrating with respect to $\mu$ and fixing a cutoff for $X^+$ such that $\mu_{\ast} = e^{\beta X^+_{\ast}}$, we obtain:
\begin{equation}
    Z_1 = - \frac{\log (\mu_E /\mu_{\ast})}{\beta} \widehat{Z}\, ,
\end{equation}
where $\widehat{Z}$ is the partition function for the 2d theory without the tachyon insertion.  Hence the partition function $Z$ in the presence of the tachyon background related to that of the theory without the tachyon $\hat Z$ via the factor $\frac{\log (\mu_E /\mu_{\ast})}{\beta} $, which thus provides an effective ``size" of the direction $X^{+}$, which matches that of \eqref{range-x+}.

The interpretation of the exponential tachyon as a wall of nothing receives further support from the dimension quenching mechanism in \cite{Hellerman:2006ff}. In  Appendix \ref{app:quench} we review it from the perspective of dynamical cobordisms in the spacetime perspective, to be discussed next.

\subsection{Spacetime description and lightlike ETW brane}
\label{sec:lightlike-spacetime}

In this section we study the spacetime description of the wall of nothing corresponding to the lightlike tachyon, and show that it satisfies the properties of (the lightlike version of) and ETW brane. This nicely confirms the worldsheet arguments of the previous section.

\subsubsection{Effective action}

In order to describe the spacetime dynamics of the lightlike tachyon configuration, we need an effective spacetime action for the relevant fields, in particular for the tachyon. This is already a subtle point, since tachyon condensation processes may in principle backreact on the whole tower of stringy states, hence the validity of the truncation to an effective theory with a finite set of fields is to some extent questionable.

In any event, this approach has been successful enough in open string tachyon effective actions, and we may venture into its use for the closed case, hoping that fortune favors the brave. 

The construction of the most general 2-derivative effective action for the metric, dilaton and tachyon in supercritical string theory has been discussed in \cite{Hellerman:2006nx} and \cite{Swanson:2008dt}, whose discussion we follow. In the string frame it has the structure
\beqa
S = \frac{1}{2}\int d^D x \sqrt{-G_{(s)}}\,e^{-2\Phi}\, \Big[
	f_1 R_{(s)} +4f_2 (\nabla \Phi)^2 - f_3 (\nabla { T})^2
	- 2f_4 - f_5 \nabla { T} \cdot \nabla \Phi \Big]\ .\quad
\label{stringframe}
\eeqa
where the $f_i(T)$ are general functions of the tachyon. By demanding that the equations of motion are compatible with the linear dilaton background with an exponential tachyon profile, one can show that the $f_i(T)$ can be expressed in terms of $f_1(T)$:
\beqa
f_2(T) & = & f_1(T) \quad ,\quad
f_3(T)  =  -f_1''(T) - \frac{f_1'(T)}{T} \quad , \quad
f_5(T)  =  4\, f_1'(T) \ ,
\nonumber \\
f_4 & = & \frac{1}{2} \biggl[ 
	4 \,f_1(T)\, \left(\frac{D-26}{6\alpha'}\right)
	+ T\,f_1'(T)\,\left( \b\cdot\b + \frac{8}{\alpha'}\right)
	- T^2\, f_1''(T)\,\b\cdot\b 
	\biggr] \ . 
\label{prepot}
\eeqa
For general exponential tachyon profiles, the tachyon background is only a solution at linearized order, hence we expect the above relations to receive corrections. For lightlike tachyons, however, the solution is exact in worldsheet perturbation theory, hence the above relations hold, and the corrections at most modify the behaviour of $f_1$ at large $T$. Note also that for lightlike tachyon profiles $\beta\cdot\beta=0$ and the tachyon potential $V_{(s)}=f_4$ becomes $\beta$-independent.

Going to the Einstein frame, we redefining the metric to absorb the $f_1$ prefactor as well as the usual dilaton factor, 
\beqa
(G_{(s)})_{MN}= e^{ \frac{
4}{D-2}\Phi} f_1^{-\frac 2{D-2}}\, G_{MN}\, ,
\eeqa
the spacetime action is
\beqa	\label{tachyon-effective-action}
S & = & \frac{1}{2}\int d^D x\sqrt{-G}\bigg[ R   - \frac{4}{D-2}\big( \partial_M \Phi\partial ^M \Phi-\frac{f_1'}{f_1}\partial_M \Phi \partial^M T\big) \\
&&
-\Big[\frac{D-1}{D-2}\frac{f_1'^2}{f_1^2}-\frac{f_1''}{f_1} - \frac{f_1'}{f_1 T }\Big]\partial_M T \partial^M T	- \frac{2}{3 \alpha'} e^{\frac{4\Phi}{D-2}}f_1^{\frac{-D}{D-2}}((D-26)f_1 + 12 T f_1') \bigg].\nonumber
\eeqa

The complete expression for $f_1$ is actually not known, beyond its expansion around $T=0$
\beqa
f_1=1-T^2+\ldots 
\eeqa
Nevertheless, \cite{Swanson:2008dt} proposed a set of regularity conditions on the effective action, which to some extent constrain $f_1$ further, and several explicit solutions were proposed, concretely $f_1=\exp (-T^2)$ and $f_1=1/\cosh(\sqrt{2}T)$. Interestingly, in the large $T$ regime (which is near the wall of nothing, our main focus), both can be parametrized as
\beqa
f_1\sim A\exp\big({\,-b \,T^k\,}\big)\, .
\label{choice-f1}
\eeqa
Actually, an outcome of \cite{Swanson:2008dt} is that the behaviour of the system is not particularly sensitive to the precise form $f_1$. In the following we focus on the dependence \eqref{choice-f1}, but later show that the same results hold, even at quantitative level, for very general forms of $f_1$.

\subsubsection{The local scalings}

We propose that the lightlike tachyon background, in the weak coupling regime, corresponds to a dynamical cobordism in $X^+$, and that the tachyon wall corresponds to an ETW brane, namely a singularity in effective theory at finite spacetime distance, and at which some scalar runs of to infinite field theory distance. In the following we show that the scalings derived from the Einstein frame spacetime solution are indeed of the ETW kind.

Note that, because the dynamical cobordism takes place via dependence on the lightlike coordinate $X^+$, in order to discuss spacetime {\em distance}, we choose slices of constant $X^0$, and measure spatial distance along $X^1$, along which the dilaton remains constant.

Again, recall that we focus on the dependence \eqref{choice-f1}, but similar conclusions hold for a very general class of profiles of $f_1$. The running scalar along $X^1$ is only the tachyon, hence the distance in field space as we approach the wall is given by:
\begin{equation}
{\cal D}=\int^{T_{\rm ETW}} \bigg(
\sqrt{\frac{f_1''}{f_1} + \frac{f_1'}{f_1}\left( \frac{1}{T} - \frac{D-1}{D-2}\frac{f_1'}{f_1}	\right)\bigg) } dT 
\sim \frac{b}{\sqrt{D-2}}\, T_{\rm ETW}^{\,k}\, ,
\end{equation}
which diverges (for $k>1$) since the tachyon goes to infinity at the ETW brane at $X^1\to \infty$.

Let us now check that the wall is indeed at finite spacetime distance in the Einstein frame. The length along $X^1$ is
\beqa\begin{aligned}
\Delta  = & \int^{\rm ETW} f_1^{\frac{1}{D-2}}\, dx^1=A^{\frac{1}{D-2}}\int^{\rm ETW} \, \exp\big(-\frac{bT^k}{D-2}\,\big)\, dx^1 
\\  = &\frac{\sqrt{2}}{bk}\,A^{\frac{1}{D-2}}\int^{\rm ETW}\exp \Big(-\frac{\cal D}{\sqrt{D-2}}\,\Big)\, \frac{d\,{\cal D}}{\cal D}= \frac{\sqrt{2}}{bk}\,A^{\frac{1}{D-2}}\, {\rm Ei}\Big(-\frac{\cal D}{\sqrt{D-2}}\,\Big)\, ,
\end{aligned}\label{lightlike-tachyon-spacetime-distance}
\eeqa
where this last function is the exponential integral, and is clearly convergent, showing that the tachyon background behaves as dynamical cobordism ending at an ETW brane, where the (tachyon) scalar runs off to infinite field space distance at a finite spacetime distance.

We can check the scaling relations of ETW branes of Section \ref{sec:dyncob}. We can expand \eqref{lightlike-tachyon-spacetime-distance} for ${\cal D}\to\infty$ as
\beqa
{\rm Ei}\Big(-\frac{\cal D}a\,\Big)\,\sim\, e^{-{\cal D}/a}\,\Big(\, \frac{a}{\cal D}+\ldots \Big)\, ,
\eeqa
and get
\beqa
\Delta\sim \exp\Big[-\frac{\cal D}{\sqrt{D-2}}-\log\Big(\frac{\cal D}{\sqrt{D-2}}\Big)
\Big]\, .
\label{distance-scaling-expf1}
\eeqa
Comparing this with \eqref{etw-scalings} gives a value 
\beqa
\delta=\frac{2}{\sqrt{D-2}}
\label{lightlike-tachyon-delta}\, .
\eeqa
Namely, we recover an exponential relation. It is interesting to point out that the log correction is reminiscent of that encountered in \cite{Buratti:2021fiv,Angius:2022aeq} for the EFT strings in \cite{Lanza:2021qsu}. Also note that, restricting to the leading exponential scaling, the critical exponent is independent of $k$. This is a particular case of the claimed robustness of the results under changes of $f_1$, and will be explored in general in Section \ref{sec:general-f1}. 

We can also compute the scaling of the Ricci scalar, which, upon direct computation gives
\beqa
|R|\sim \exp\Big[\frac{2\cal D}{\sqrt{D-2}}+\log\Big(\frac{{\cal D}^2}{D-2}\Big)
\Big]\,.
\eeqa
Again, the leading terms gives the scaling corresponding to an ETW brane, with $\delta$ given by \eqref{lightlike-tachyon-delta}, again remarkably independent of $k$.\\
The potential, computed in the limit $T \mapsto \infty$ with \eqref{tachyon-effective-action}, also agrees nicely with the general formula provided by the local analysis in \eqref{eq:sol-V}:
\begin{equation}
    V(T)=- \frac{8}{ \alpha'} A^{- \frac{2}{D-2}} k b \,T^k e^{\frac{4 \phi}{D-2}} e^{\delta \mathcal{D}} = -a \,c(T)\, e^{\delta \mathcal{D}}\, , 
\end{equation}
with $a \in \left[ 0,1 \right]$ and the subleading polynomial correction can be absorbed by the function $c(T)$\footnote{More details about such subleading corrections can be found in the Appendix B of \cite{Angius:2022aeq}.}.

Even more remarkably, the value of $\delta$ \eqref{lightlike-tachyon-delta} for the lightlike tachyon agrees with the critical exponent \eqref{linear-dilaton-delta} of the ETW brane at the beginning of time of the linear dilaton solution. This shows that both kinds of ETW branes are very similar, and  is strongly suggestive that they may admit similar microscopic descriptions.  Hence, we claim that the singularity at the beginning of time is a dynamical cobordism to nothing triggered by (the strong coupling version of) the condensation of the closed string tachyon.
We look deeper into this argument in Section \ref{sec:recombination} but before then, we show that this surprising matching of the critical exponents holds for general profiles of $f_1$.

\subsubsection{General $f_1$}
\label{sec:general-f1}

Let us now show that the above structure, and in particular the same value for the critical exponent $\delta$, holds for general $f_1$ under very mild conditions. In particular we demand that $f_1$ decays at large $T$ faster then $1/T$. This is a very reasonable requirement, in particular notice that this ensures the convergence of the integral for the spacetime distance $\Delta$ to the ETW brane. Hence it implements the intuition that the wall of nothing propagating at the spped of light hits in finite time any point at finite spacetime distance.

Consider now the integral for the field space distance
\beqa
{\cal D} &=& \int \left[\frac{f_1''}{f_1} + \frac{f_1'}{f_1}\left( 
		\frac{1}{T} 
		- \frac{D-1}{D-2}\frac{f_1'}{f_1}
		\right) 
		\right] ^{\frac 12} \, dT \,.
\eeqa
We start massaging the integrand of ${\cal D}$, by noticing that
\beqa
\frac{f_1''}{f_1} + \frac{f_1'}{f_1}\left( 
		\frac{1}{T} 
		- \frac{D-1}{D-2}\frac{f_1'}{f_1}
		\right) \,=\,  \left(\frac{f_1'}{f_1}\right)'+\frac{f_1'}{T\,f_1}+\frac{1}{D-2}\left(\frac{f_1'}{f_1}\right)^2\,,
		\label{before-the-approx}
\eeqa
it is easy to show that for $f_1$ decaying faster than $1/T$, the dominant term is the last one. In fact one can see by considering different profiles (e.g. power-law, exponential, exponential of an exponential, etc) that, the faster the decay, the more the last terms dominates. Then
\beqa
{\cal D} \sim\int \frac{1}{\sqrt{D-2}} \frac{f_1'}{f_1}\, dT\,.
\eeqa
We may write this as
\beqa
d{\cal D}=\frac{1}{\sqrt{D-2}} \frac{f_1'}{f_1}\, dT=\frac{1}{\sqrt{D-2}} d\log f_1=\sqrt{D-2}\,\, d\log f_1^{\frac 1{D-2}}\,.
\eeqa
Namely
\beqa \label{eq:f1andD}
f_1^{\frac 1{D-2}}=\exp(-{\cal D}/\sqrt{D-2})\,,
\eeqa
where we have chosen the appropriate sign for the distance to be positive (recall that $f_1$ is a function that decreases to zero). This allow to express the spacetime distance as
\beqa\label{eq:startingpoint}
\Delta\sim\int f_1^{1/(D-2)} dx^1\sim\int \exp(-{\cal D}/\sqrt{D-2}) \frac{dT}T\, .
\eeqa
This has a similar structure to the intermediate expression in
\eqref{lightlike-tachyon-spacetime-distance}. Similar to the exponential integral there, the above integral behaves just like the exponential in the integrand, leading to the scaling 
\beqa \label{eq:scalingrel}
\Delta=\exp(-{\cal D}/\sqrt{D-2})\, ,
\eeqa
which reproduces the value of $\delta$ in \eqref{lightlike-tachyon-delta}. Indeed, one can check that the additional terms in the integrand lead to subleading corrections, of the kind in \eqref{distance-scaling-expf1} (for a proof of this statement under mild assumptions, we refer the reader to Appendix \ref{app:scalingrelation}). 

\section{The strong coupling region and the origin of time}
\label{sec:recombination}

In this section we argue that the microscopic description of the ETW brane at the beginning of time is a region of (the strong coupling version of) closed string tachyon condensation.

\subsection*{The ETW brane recombination}

Let us now consider the full lightlike tachyon configuration, including the strongly coupled region, and consider the interplay of the two ETW branes we have encountered.

In the string frame variables there are two asymptotic regions, controlled by seemingly different physics.
The first is the region $X^0\to -\infty$, with $X^1$ finite (hence $X^+\to -\infty$), which corresponds to a linear timelike dilaton configuration, with negligible tachyon background. The second is the region $X^+\to \infty$ at finite $X^1$ (hence $X^0\to\infty$), which corresponds to a lightlike tachyon configuration at weak string coupling.  Both regions are disjoint, as they only coincide at infinity in $X^0\to -\infty$, $X^+\to\infty$ (hence we need $X^1\to\infty)$. 

In the Einstein frame, these asymptotic regimes turn into singularities at finite distance in spacetime, triggered by the running off of suitable scalars (the tachyon or the dilaton) to infinite distance in field space. Following the dynamical cobordism interpretation advocated in \cite{Buratti:2021yia,Buratti:2021fiv,Angius:2022aeq}, these are ETW branes chopping off the region of spacetime beyond them. 

An important observation is that the effective theory in which one describes ETW branes is not valid at arbitrarily short distances to the singularity. The singularity is expected to be smoothed out by new UV physics which implies the existence of a cutoff in the applicability of the effective theory. This translates into cutting of a strip of spacetime around the singularities, hence providing a notion of `strechted' ETW brane in effective theory. This can be obtained in different ways, for instance by imposing a maximal bound on the scalar curvature. Instead, we use a criterion directly inspired by the swampland distance conjecture \cite{Ooguri:2006in}, as follows.

The distance conjecture states that when an effective theory reaches to a large distance ${\cal D}$ in field space, its effective cutoff scales as 
\beqa
\Lambda\sim e^{-\alpha{\cal D}}\, ,
\label{sdc-cutoff}
\eeqa
for some order 1 coefficient $\alpha$. The actual distance conjecture moreover claims that there is an infinite tower of states becoming light with $\Lambda$, but this formulation corresponds to an adiabatic motion in moduli space, and such towers may actually not arise in dynamical situations with spacetime dependence of the scalars \cite{Buratti:2018xjt}\footnote{A simple example is the Taub-NUT geometry, which can be regarded as a spacetime-dependent solution of $\IS^1$ compactifications, in which the circle shrinks to zero size at a point in the base. Hence, it attains infinite distance in the naive circle compactification moduli space, but no tower of light particles or other disasters arise.}. Hence we stick to the milder statement that a cutoff is developed, whose origin in our context would stem from the UV completion of the ETW brane.

In our configuration we hence consider the slice of spacetime at which the field space distance (in the combined  tachyon-dilaton system) reaches a large but finite value. From the Einstein frame action \eqref{tachyon-effective-action}, the relevant kinetic terms read
\begin{equation}
\frac{4}{D-2}\partial\Phi\cdot\partial\Phi\,-\, \frac{4}{D-2}\frac {f_1'}{f_1}\partial\Phi\cdot\partial T\, +\, \left(\,-\frac{f_1''}{f_1}-\frac{f_1'}{Tf_1}+\frac{D-1}{D-2} \frac{f_1'{}^2}{f_1^{\,2}}\right) \partial T\cdot\partial T\,.
\end{equation}
We are interested in the behaviour near the intersection of the two singularities. Since this lies at large $T$, we can simplify the last term using the argument in section \ref{sec:general-f1}. Using $(f_1'/f_1)\partial T=\partial \log f_1$, the kinetic term may be written
\beqa
\frac{1}{D-2}(\,2\partial\Phi-\partial \log f_1\,)^2\, ,
\eeqa
so that the slices of constant distance are defined by
\beqa
{\cal D}\sim \frac{1}{\sqrt{D-2}}(2\Phi-\log f_1)={\rm const}\, .
\label{constant-distance-slice}
\eeqa
Note that interestingly, the swampland distance cutoff \eqref{sdc-cutoff} is
\beqa
\Lambda\sim e^{-\alpha {\cal D}}\sim \exp \big(-\frac{\alpha \delta}{2} \,(2\Phi-\log f_1)\big)=\big(e^{-2\Phi}f_1\big)^{\alpha\delta/2}\, ,
\eeqa
where $\delta$ is given by \eqref{lightlike-tachyon-delta}. The factor inside brackets is the prefactor of the Einstein term in the string frame action. The fact that it relates to the cutoff scale shows that one gets the same spacetime slices if one uses a bound in the scalar curvature to limit the applicability of the effective theory, rather than in the field space distance. Indeed, this is expected from the scaling \eqref{scalings-phi-sigma-R} of $R$ with ${\cal D}$ near ETW branes.

\begin{figure}[htb]
\begin{center}
\includegraphics[scale=.4]{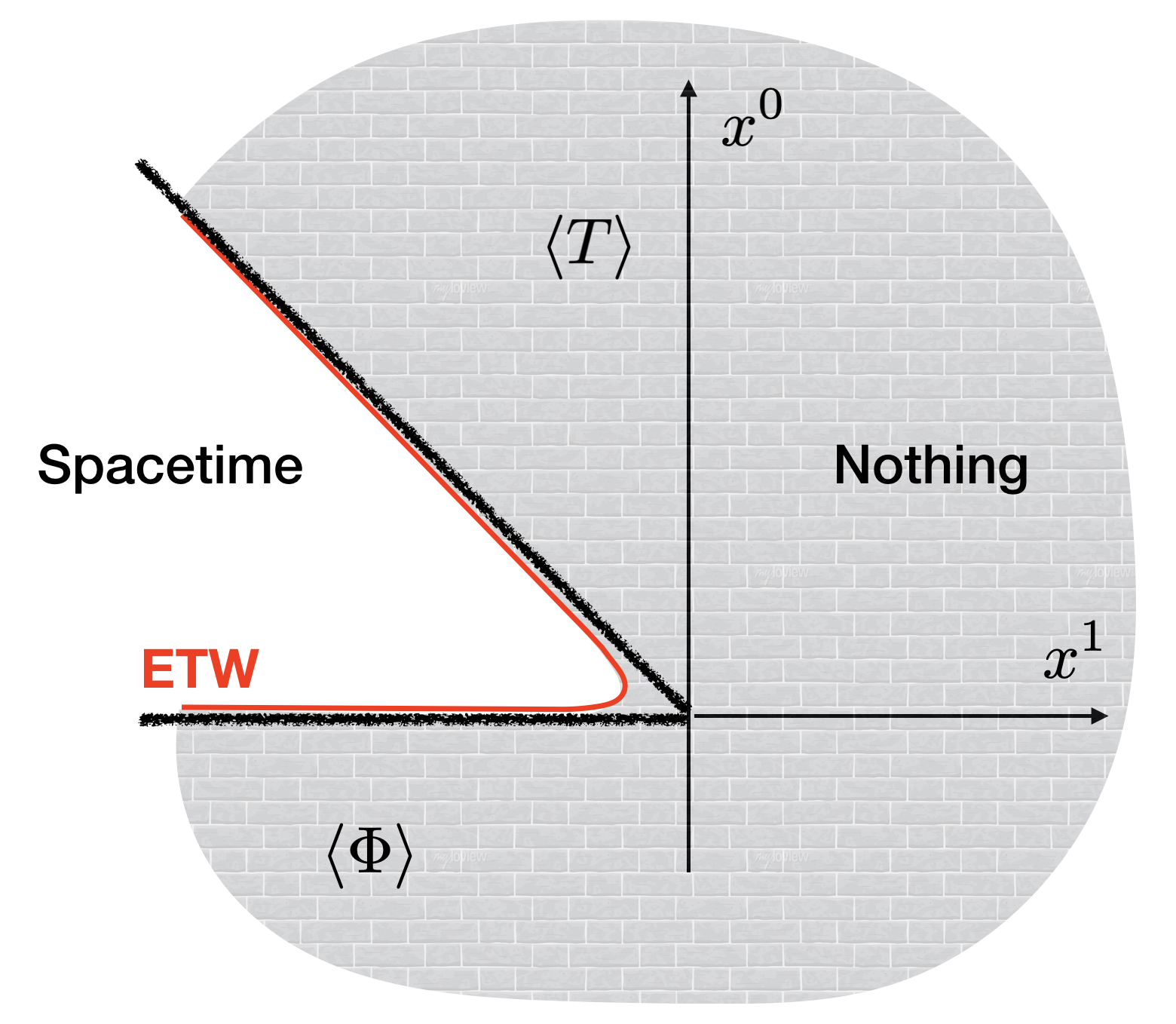}
\caption{\small Depiction of a spacetime slice at which the solutions attain a large fixed field space distance in the tachyon-dilaton space. The asymptotes near $x^0\sim 0$ and $x^+\sim 0$, should be regarded as corresponding to the physical location of the ETW branes, which are hence recombined in the middle region, and which for a boundary of spacetime, with nothing beyond them. The spray line signal the location of the singularities.}
\label{fig:slice}
\end{center}
\end{figure}

The curve in the $(x^0,x^1)$-plane defined by \eqref{constant-distance-slice} asymptotes to constant $X^0$ on one side and to constant $X^+$ on the other. For illustration we may consider $f_1$ as in \eqref{choice-f1} and get
\begin{equation}
2qx^0-b\mu^k e^{\beta k x^+}={\rm const.}
\end{equation}
This leads to slices of the form
\beqa
x^1=\frac{\sqrt{2}}{k\beta}\log(x^0+cst)-x^0+\frac{\sqrt{2}}{k\beta}\log \frac{2q}{b\mu^k}\, ,
\eeqa
for some constant $cst$ related to the cutoff. In Figure \ref{fig:slice} we depict the structure of such curves and of the resulting spacetime picture.

From this it is clear that what seemed to be an intersection between the two ETW walls is in fact a smooth region that interpolates between the two. This hence strongly motivates that this solution describes a one and only recombined ETW wall. Another interesting indication for this is the fact that the dilaton-tachyon mixing in the effective action (scaling like $f_1'/f_1$ times powers of the string couplings) gets large for large tachyon and strong coupling, namely near the naive intersection of the singularities. Hence, the two walls, which asymptotically correspond to the dilaton or the tachyon running off to infinite distance in their field space, become of a very similar nature in that intersection region.
\\

As a side note, one may wonder if our analysis extends to the case of critical $D=26$ bosonic string theory. Indeed, this theory is tachyonic and, as mentioned in section \ref{sec:linear-dilaton}, it admits a lightlike dilaton background. It is then clear from the equation of motion for the tachyon \eqref{tachyon-eom} that, if we take the dilaton along the direction $X^{+}$, the only option is to take the tachyon along $X^{-}$. This means that the ETW branes corresponding to large tachyon and strong coupling are of the intersecting kind, exactly like those described in this section. It would be interesting to explore these lightlike linear dilaton backgrounds in other critical theories, and make contact with existing proposals concerning the resulting singularities \cite{Craps:2005wd,Hellerman:2006ff}.

\medskip

\section{Conclusions}
\label{sec:conclusions}

In this paper we have studied timelike linear dilaton backgrounds of supercritical string theories as time-dependent solutions in string theory, and addressed the question of the resulting spacelike singularity, from the perspective of the cobordism conjecture. We have quantitatively characterized the solution as a Dynamical Cobordism in which the dilaton rolls until it hits infinite field space distance at a singularity at finite time in the past. We have shown that the singularity in effective theory follows the scaling behaviour of ETW branes. In order to clarify its microscopic description, we have considered lightlike tachyon condensation backgrounds, whose microscopic description had been argued to correspond to a stringy version of a bubble of nothing. Using an effective theory approach, we have characterized them as ETW branes and have encountered precisely the same scaling exponent as the beginning of time singularity. Together with the fact that both ETW branes join smoothly from the effective theory perspective, this has motivated our proposal that the spacelike singularity should correspond in string theory to a region of (a strong coupling version of) closed tachyon condensation, giving rise to a cobordism boundary defining the beginning of time.

There are several open questions and new directions:
\begin{itemize}
    \item We have used an effective theory for the tachyon in terms of the undetermined function $f_1$. Even though our results are robust under changes of the precise form of this function, it would be interesting to determine it, or at least its asymptotic behavior for large $T$. 
    \item Conversely, it would be interesting to understand if the criterion that the theory allows for a resolution of spacelike singularities can be used as a constraint on effective theories. For instance, there exist choices of $f_1$ which lead to lightlike tachyon ETW branes with scalings different from the beginning of time one. Such theories may not be compatible with a microscopic description of the latter, as the ETW brane may not be compatible for recombination. It is thus tantalizing to claim that this can be used as a criterion to exclude such choices of $f_1$. It would also be interesting to understand this possibility, possibly invoking other swampland constraints or physical considerations.
    \item Although we have focused on the bosonic theory, there is a rich set of phenomena arising in lightlike tachyon backgrounds in other string theories. We expect our ideas to lead to interesting new insights into this web of transitions.
    \item Finally, an interesting corner in this circle of ideas is that of lightlike dilaton backgrounds in critical string theories. They are toy models of cosmological singularities, which in certain supersymmetric cases admit interesting proposals for their microscopic description \cite{Craps:2005wd}. It would be exciting to use cobordism ideas to make progress on the understanding of such backgrounds.
\end{itemize}
We hope to report on these and other questions in the near future.

\medskip


%
\section*{Acknowledgments}

We are pleased to thank Luis Ib\'anez, Fernando Marchesano, Miguel Montero, Lorenz Schlechter and Irene Valenzuela for useful discussions, and Ginevra Buratti, Jos\'e \- Calder\'on, Jes\'us Huertas for collaboration on related topics. M.D. also wishes to acknowledge the hospitality of the Department of Physics of Harvard University during the development of this work. This work is supported through the grants CEX2020-001007-S and PGC2018-095976-B-C21, funded by MCIN/AEI/10.13039/501100011033 and by ERDF A way of making Europe. The work by R.A. is supported by the grant BESST-VACUA of CSIC.  The work by M.D. is supported by the FPI gran no. FPI SEV-2016-0597-19-3 from Spanish National Research Agency from the Ministry of Science and Innovation.

\newpage

\appendix
\section{Dimension quenching as an interpolating domain wall}
\label{app:quench}

In \cite{Buratti:2021fiv} it was argued that, when the scalars remain at finite distance in field space as one hits the wall, the corresponding configuration described an interpolating wall between different QG theories. This scenario is built in contrast with the end-of-the-world walls where the fields reach infinite distance in field-space at the wall. Instead of the solution ending abruptly at the location of a singularity, these interpolating solutions continue across the wall into another theory. On each side of the interpolating wall, the field spaces may have different structures but the location of the wall itself is at finite distance in both of them. As a result, the interpolating wall must have all the right properties for communicating between the two theories, whatever they may be. It is clear from this that the microscopic nature of these walls can be hard to describe; it may in general be non-supersymmetric and may involve strong-coupling physics. The existence of such objects is one of the predictions of the Cobordism Conjecture  \cite{McNamara:2019rup}.  \\
The examples of interpolating walls in \cite{Buratti:2021fiv} were simple enough to be described by standard supersymmetric objects. For example, D8 branes in massive type IIA string theory were identified as such walls interpolating between ``different'' massive IIA theories with different units of 0-form flux. Here, we re-interpret the dimension-quenching bubbles of \cite{Hellerman:2006ff,Hellerman:2007fc} as interpolating walls between bosonic string theories of different dimensions. 

Throughout this paper we have dealt with tachyons with exponential profiles along one light-like direction. These solutions were shown to lead to bubbles of nothing, which fit in the wall-of-nothing description. There are, however, slightly more complicated solutions to the tachyon equation of motion \eqref{tachyon-eom}. Following \cite{Hellerman:2006ff}, we can consider a profile with oscillatory dependence on another coordinate, denoted by $X^2$: 
\beqa
T(X) = \mu_0^2 \exp ( \beta X^{+}) - \mu_ k^2 \cos ( k X^2) \exp( \beta_k X^{+})\, .
\eeqa
This is a solution to the equation of motion with a timelike linear dilaton $\Phi= -q X^0$ background if: 
\beqa  q \beta_k = \sqrt{2}\,\bigg(\,\frac{2}{\alpha'}- \frac{1}{2}k^2\,\bigg)\,.
\eeqa
Since the tachyon couples to the worldsheet as a potential, the theory has a vacuum at $X^2 = 0$. One can show that expanding around this vacuum in the limit where the wavelength of oscillations $k^{-1}$ is much larger than the string length $l_s$ yields:
\beqa \label{eq:quench-profile-exp}
T(X^{+},X^2)= \frac{\mu^2}{2\alpha'}\exp ( \beta X^{+}) : (X^2)^2: \, ,
\eeqa
where $\mu^2 = \alpha'k^2 \mu_k^2$, and dots denote normal ordering. We refer the reader to \cite{Hellerman:2006ff} for additional aspects of the detailed derivation.

The physical interpretation of this is clear. Before the tachyon condenses, at $X^{+}\to - \infty$, the string propagates in $D-1$ spatial dimensions. As the string reaches a regime where $T\sim 1$ (namely $X^+\sim \beta^{-1}\log \mu$), the potential confines the string to the region where it is vanishing, at $X^2= 0$. Strings that oscillate along the $X^2$ dimension will be expelled from the region of large tachyon condensate. This bubble thus interpolates between a region of $D-1$ spatial dimensions to one where the string can effectively propagate in $D-2$ dimensions. These types of bubbles were dubbed dimension-quenching or dimension-changing bubbles. 

Turning now to the dynamical cobordism perspective, one can see in \eqref{eq:quench-profile-exp} that the tachyon field remains at a finite value (hence at a finite distance in field space) at the location of this bubble at $X^2=0$. This fits perfectly with the description of an interpolating wall as described in \cite{Buratti:2021fiv}. We thus interpret these dimension-quenching bubbles as examples of dynamical cobordism interpolating walls between bosonic theories of different dimensions.

As a side note, one can construct similar bubbles that kill more than one dimension by granting oscillatory dependence of the tachyon on extra dimensions. Furthermore, this dimension-quenching mechanism also extends to superstring theories and can be used to draw connections between supercritical Type 0 theories and their 10-dimensional critical counterparts \cite{Hellerman:2006ff}.

\section{The partition function with a lightlike tachyon background}
\label{app:partitionfunction}

The computation of the partition function in the presence of the lightlike tachyon background is obtained evaluating the path integral without vertex operator insertions:
\begin{equation}    \label{partitionfunction}
    Z \left( \mu \right) = \int \mathcal{D} X^+ \mathcal{D} X^- \mathcal{D}X^i \mathcal{D}g \mathcal{D}(ghosts) e^{i S^{deformed}} \, ,
\end{equation}
where we have emphasized the fact that the integration along the lightlike directions does not affect the spacelike directions ($i=2,...,D-1$).  Being at weak coupling, we only consider the one-loop contribution and so we have to evaluate the 2d action on a genus one worldsheet:
\begin{equation}
\begin{split}
 S^{deformed} = & - \frac{1}{2 \pi \alpha'} \int_{\mathcal{M}_1} d^2 \sigma \sqrt{g}  g^{\alpha \beta} \left[ -\partial_{\alpha} X^+ \partial_{\beta} X^- - \partial_{\alpha} X^- \partial_{\beta} X^+ + \partial_{\alpha} X^i \partial_{\beta} X_i  \right]  + \\
& + \frac{1}{2 \pi} \int_{\mathcal{M}_1} d^2 \sigma \sqrt{g} R_2 \phi (X) - \frac{1}{2 \pi} \mu \int_{\mathcal{M}_1} d^2 \sigma \sqrt{g} e^{\beta X^+} \, . 
\end{split}
\end{equation}
In analogy with the procedure in \cite{McGreevy:2005ci}, we decompose the field $X^+(\tau, \sigma)$ into its zero and nonzero modes:
\begin{equation}
X^+ (\tau, \sigma) = X^+_0 + \widehat{X}^+(\tau, \sigma)\, ,
\end{equation}
we get a standard integration for the zero mode:
\begin{equation}
\mathcal{D}X^+ = dX^+_0 \mathcal{D} \widehat{X}^+.
\end{equation}
Choosing the following convention to perform a Wick rotation:
\begin{equation}
    \tau \mapsto i \tau_E \quad \quad \quad X^i \mapsto i X^i_E \quad \quad \quad \mu \mapsto -i \mu_E \, ,
\end{equation}
equation \eqref{partitionfunction} becomes:
\begin{equation}
Z \left( \mu \right) = \int dX^+_0 \int \mathcal{D} \widehat{X}^+ \mathcal{D} X^- \mathcal{D}X^i \mathcal{D}g \mathcal{D}(ghosts) e^{- S^{deformed}_{E}}\, .
\end{equation}
where the tachyonic potential gives an oscillating contribution to the integral in the condensate region. Such a behavior produces a truncation of the contributions to the integral coming from configurations with $X^+ \mapsto \infty$. The Euclidean 2d action is:
\begin{equation}\label{EuclidianAction}
\begin{split}
    S_E^{deformed} = & \frac{1}{2 \pi \alpha'} \int d^2 \sigma_E \sqrt{g} \left[\partial_{\sigma^0} \widehat{X}^+ \partial_{\sigma^0} X^- + \partial_{\sigma^1} X^- \partial_{\sigma^1} \widehat{X}^+ +  \partial_{\alpha} X^i \partial_{\alpha} X^i \right] \\  & + \frac{1}{2 \pi} \int d^2 \sigma_E \sqrt{g} R_2 \phi (X) + \frac{i \mu_E}{2 \pi} \int d^2 \sigma_E \sqrt{g} e^{\beta X^+} \, .\\ \end{split}\end{equation}
Using the variable change:
\begin{equation}y= e^{\beta X^+_0} \quad \quad \quad  \longrightarrow \quad \quad \quad dX^+_0 = \frac{dy}{\beta y},\end{equation}
and making the dependence of the integrand on $X^{+}_0$ explicit, we obtain:
\begin{equation}
Z (\mu_E) = \int \mathcal{D} \widehat{X}^+ \mathcal{D} X^-  \mathcal{D}X^i \mathcal{D}g \mathcal{D}(ghosts) \int_0^{\infty} \frac{dy}{\beta y} e^{-S_E^{kinetic}- S_E^{dilaton} - \frac{i \mu_E}{2 \pi} \int d^2 \sigma_E \sqrt{g} y e^{\beta \widehat{X}^+}} \, ,
\end{equation}
where $S^{kinetic}_E+S^{dilaton}_E = S^{free}_E$ are respectively the kinetic and the dilaton contributions in the Euclidean action \eqref{EuclidianAction}. \\
Now, let us consider the following quantity:
\begin{equation}
    \frac{\partial Z}{\partial \mu_E} = \int \mathcal{D} \widehat{X}^+ \mathcal{D} \textit{(others)} \int_0^{\infty} \frac{dy}{\beta} e^{-S_E^{free}+ \frac{-i \mu_E}{2 \pi} \int d^2 \sigma_E \sqrt{g} y e^{\beta \widehat{X}^+}} \left(  \frac{-i}{2 \pi} \int d^2 \sigma_E \sqrt{g} e^{\beta \widehat{X}^+} \right)\,.
\end{equation}
Let us finally perform the integration on the zero mode. We obtain:
\begin{equation}
      \frac{\partial Z}{\partial \mu_E} = -\frac{1}{\beta \mu_E} \int \mathcal{D} \widehat{X}^+ \mathcal{D} X^- \mathcal{D}X^i \mathcal{D}g \mathcal{D}(ghosts) e^{-S_E^{free}} \, ,
\end{equation}
where $S_E^{free}$ is the euclidean action of the world-sheet theory without the tachyon potential. Integrating with respect to $\mu$ and fixing a cutoff for $X^+$ such that $\mu_{\ast} = e^{\beta X^+_{\ast}}$, we obtain:
\begin{equation}
    Z_1 = - \frac{\log (\mu_E /\mu_{\ast})}{\beta} \widehat{Z}\, ,
\end{equation}
where $\widehat{Z}$ is the partition function for the free 2d theory, namely without the tachyon insertion and without integrating the zero modes of $X^+$. 
Note that the tachyon's contribution to the partition function is entirely encoded in the zero modes. \\
We can interpret this factor as a ``size" of the direction $X^{+}$. Indeed, because of the potential barrier created by the condensation of the tachyon, no physical degrees of freedom penetrate inside the bubble wall, beyond $X^{+}\sim 1$. As mentioned previously, the path integral is suppressed in this region. The direction $X^{+}$ thus has an effectively finite ``size'' that agrees with the estimate in  \eqref{range-x+}. 

\section{The critical exponent for general $f_1$}
\label{app:scalingrelation}

In this appendix we provide more details regarding how we obtain the scaling relation \eqref{eq:scalingrel} for a general $f_1$ decaying faster than $T^{-1}$. The starting point is \eqref{eq:startingpoint}, which we rewrite as follows: \beqa\Delta\sim \int \exp \big(-\frac{\cal D}{\sqrt{D-2}} - \log (T\mathcal{D}')\;\big) d\mathcal{D}\, ,\eeqa
where the prime stands for derivation with respect to $T$. Proving that the first term in the exponential is the dominant one in the limit $T\to \infty$ comes down to comparing the two terms:
\begin{equation}
 -\frac{\cal D}{\sqrt{D-2}}\sim \log f_1\;\;\;\text{and}\;\;\;-\log T \mathcal{D}' \sim \log \Big( \frac{f_1}{| f_1'| T }\Big)\,.   
\end{equation}
Notice that in the special case where $f_1$ is power-like $f_1= T^{-k}$, with $k>0$, the second term in the exponential is constant so one automatically obtains the scaling relation \eqref{eq:scalingrel}. For other choices of $f_1$, we wish to check that in the limit $T\to \infty$,
\begin{equation}\label{eq:dominancecondition}
    \frac{| \log ( \frac{f_1}{| f_1'| T })|}{| \log f_1|} \to 0\,.
\end{equation}
We consider a positive and monotonically decreasing function $f_1$, we require: 
\begin{equation}\label{eq:conditionf1}
    \Big|\log \Big( \frac{f_1}{| f_1'| T }\Big)\Big| << |\log f_1| \;\text{ as}\;\; T \to \infty\,.
\end{equation}

We know for a fact that $\log f_1$ is negative when $T\to \infty$. As we will show shortly, $\log ( \frac{f_1}{| f_1'| T })$ is negative, then one can easily show that \eqref{eq:conditionf1} implies:
\begin{equation}
 |f_1'| << T^{-1} \, ,
\end{equation}
which is true for any function $f_1$ under consideration since $f_1<< T^{-1}$.

The question of whether \eqref{eq:dominancecondition} is verified is thus recast into the question of the sign of $\log ( \frac{f_1}{| f_1'| T })$. In order to determine if this term is negative, we can write it as follows,
\begin{equation}
    \log \Big( \frac{f_1}{| f_1'| T }\Big)= \log\big(\frac{f_1}{T}\big) - \log(|f_1'|)\, .
\end{equation}
One clearly sees that the first term is negative whilst the second is positive. We would therefore like to show that the first one dominates: 
\begin{equation}
    |\log(\frac{f_1}{T})| >> |\log(|f_1'|) \;\;\to\;\; \frac{f_1}{T}<< |f_1'| 
\end{equation}
This is condition is verified for all valid choices of $f_1$ that are not power-like. Indeed, we can use the trivial fact that $f_1< < T$ as $T\to \infty$ to show that  $\frac{f_1}{ |f_1'| }< <T$. In the case where $f_1$ is a power-like function the inequality is not strict as $\frac{f_1}{T}\sim |f_1'| $, but we still recover the right scaling relations as mentioned previously.

\newpage
\bibliographystyle{JHEP}
\bibliography{mybib}

\end{document}